\documentclass{article}

\usepackage{arxiv}

\usepackage[utf8]{inputenc} 
\usepackage[T1]{fontenc}    
\usepackage{hyperref}       
\usepackage{url}            
\usepackage{booktabs}       
\usepackage{amsfonts}       
\usepackage{nicefrac}       
\usepackage{microtype}      
\usepackage{lipsum}		
\usepackage{graphicx}
\usepackage{natbib}
\usepackage{doi}

\usepackage{amsfonts}       
\usepackage{nicefrac}       
\usepackage{microtype}      
\usepackage{cool}
\usepackage{colortbl}%
\usepackage{tikz}
\usetikzlibrary{positioning}

\DeclareMathAlphabet{\mathcal}{OMS}{cmsy}{b}{n}

\newcommand{\changed}[1]{{#1}}

\newcommand{\vb}{{\mathbf b}}

\newcommand{\vu}{{\mathbf u}}

\newcommand{\vv}{{\mathbf v}}
\newcommand{\vw}{{\mathbf w}}

\newcommand{\vx}{{\mathbf x}}
\newcommand{\xx}{{\mathbf x}}
\newcommand{\vy}{{\mathbf y}}

\newcommand{\vNull}{{\mathbf 0}}
\newcommand{\vxi}{{\boldsymbol \xi}}
\newcommand{\vomega}{{\boldsymbol \omega}}

\newcommand{\mB}{{\mathbf B}}

\newcommand{\mI}{{\mathbf I}}

\newcommand{\mQ}{{\mathbf Q}}

\newcommand{\mT}{{\mathbf T}}
\newcommand{\mU}{{\mathbf U}}

\newcommand{\mY}{{\mathbf Y}}

\newcommand{\Transp}{{{\mathrm T}}}

\newcommand{\cP}{{\mathcal P}}

\Style{IdentityMatrixSymb=\mI,%
       DSymb={\mathrm d},%
       IntegrateDifferentialDSymb={\mathrm d}}

\title{On the Objectivity and Quasi-Objectivity of TSE and TRA}

\date{} 					

\author{ \href{https://orcid.org/0000-0002-8009-7070}{\includegraphics[scale=0.06]{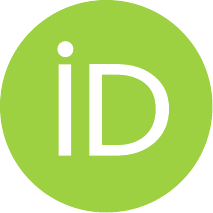}\hspace{1mm}Holger Theisel}
		\\
	Department of Computer Science\\
	University of Magdeburg\\
	Magdeburg, Germany\\
	\texttt{theisel@ovgu.de} \\
	\And
	\href{https://orcid.org/0000-0002-5469-1324}{\includegraphics[scale=0.06]{orcid.pdf}\hspace{1mm}Anke Friederici} \\
	Department of Computer Science\\
	University of Magdeburg\\
	Magdeburg, Germany\\
	\texttt{anke@isg.cs.uni-magdeburg.de} \\
	\And
	\href{https://orcid.org/0000-0002-3020-0930}{\includegraphics[scale=0.06]{orcid.pdf}\hspace{1mm}Tobias G{\"u}nther} \\
	Department of Computer Science\\
	Friedrich-Alexander-University Erlangen-N{\"u}rnberg\\
	Erlangen, Germany \\
	\texttt{tobias.guenther@fau.de} \\
}

\renewcommand{\shorttitle}{On the Objectivity and Quasi-Objectivity of TSE and TRA}

\begin{document}
\maketitle

\begin{abstract}
We analyze two recently-introduced flow measured that are based on a single trajectory only: trajectory stretching exponent (TSE) to  detect hyperbolic (stretching) behavior, and trajectory angular velocity (TRA) to detect   elliptic (rotation) behavior. \cite{Haller21singletrajecory} and ~\cite{Haller21singletrajecory_erratum} introduced TSE, TRA as well as the concept of quasi-objectivity, and formulated theorems about the objectivity and quasi-objectivity of TSE and TRA.

In this paper, we present two counter-examples showing that all theorems in \cite{Haller21singletrajecory} and ~\cite{Haller21singletrajecory_erratum} are incorrect.
\end{abstract}

\keywords{Objectivity \and Flow Analysis}

\section{Introduction}
Recently, \cite{Haller21singletrajecory} and \cite{Haller21singletrajecory_erratum}
introduced measures based on a single trajectory only.
For this, the concept of  quasi-objectivity is introduced: Contrary to classical objectivity where a scalar value must be invariant under arbitrary time-dependent Euclidian transformations, for quasi-objectivity a condition (A) is introduced, and invariance is not demanded for all Euclidean transformations but only for those fulfilling (A).  Then, \cite{Haller21singletrajecory}  introduced several measures based on a single trajectory: {\em extended trajectory stretching exponents} $\mbox{TSE}$ and 
$\overline{\mbox{TSE}}$, and {\em extended trajectory angular velocity} $\mbox{TRA}$,    $\overline{\mbox{TRA}}$. 
\cite{Haller21singletrajecory}  claimed that  $\mbox{TSE}$ and 
$\overline{\mbox{TSE}}$ are objective in the extended phase space, and that $\overline{\mbox{TRA}}$ is quasi-objective in the extended phase space under a certain condition put to the average vorticity in a certain neighborhood of the trajectory. The new measures have been applied in several follow-up papers: 
\cite{bartos22_02}, \cite{aksamit_haller_2022}.

In this paper, we show that the claims by \cite{Haller21singletrajecory} concerning objectivity of   $\mbox{TSE}$, $\overline{\mbox{TSE}}$, $\overline{\mbox{TRA}}$ are incorrect. In fact, we show by a counter-example that  neither   $\mbox{TSE}$ nor 
$\overline{\mbox{TSE}}$ are objective in the extended phase space. Further, $\overline{\mbox{TRA}}$ is not quasi-objective in the extended phase space  under an averaged-vorticity-based condition. 

In an erratum, \cite{Haller21singletrajecory_erratum} reformulate claims about quasi-objectivity of $\mbox{TSE}$, $\mbox{TRA}$. We show by a second counter-example that the new claims in the erratum \cite{Haller21singletrajecory_erratum} are incorrect as well.

In summary, our two counter-examples show that all theorems in \cite{Haller21singletrajecory} and \cite{Haller21singletrajecory_erratum} are incorrect.

We note that the first counter-example was already published in \cite{theisel22_01}, and \cite{Haller21singletrajecory_erratum} was published as a reaction on \cite{theisel22_01}.

\section{$\mbox{TSE}$, $\mbox{TRA}$, and objectivity}

Objectivity, a concept from continuum mechanics, refers to the invariance of a measure under a moving reference system. Let $s(\vx,t)$, $\vw(\vx,t)$, $\mT(\vx,t)$ be a time-dependent scalar field, vector field and tensor field, respectively. Further, let $\tilde{s}(\tilde{\vx},t)$, $\tilde{\vw}(\tilde{\vx},t)$, 
 $\tilde{\mT}(\tilde{\vx},t)$ be their observations under the Euclidean frame change
 \begin{equation}
\label{eq_define_movingsystem}
\vx = \mQ(t) \; \widetilde{\vx} + \vb(t)
\end{equation}
where $\mQ=\mQ(t)$ is a time-dependent rotation tensor and  $\vb(t)$ is a time-dependent translation vector. Then $s, \vw, \mT$ are \emph{objective} if the following conditions hold, cf. \cite{truesdell_book}:
 \begin{align}
\tilde{s}(\tilde{\vx},t)=s(\vx,t) \;,\; \tilde{\vw}(\tilde{\vx},t) = \mQ^\Transp \, \vw(\vx,t) \;,\;
\tilde{\mT}(\tilde{\vx},t) =  \mQ^\Transp \, \mT(\vx,t) \,   \mQ.
\end{align} 

\subsection{$\mbox{TSE}$ and $\mbox{TRA}$ in a nutshell}

\cite{Haller21singletrajecory} introduced measures for stretching and rotation that are based on single trajectories only: 
 {\em Extended trajectory stretching exponents} $\mbox{TSE}$ , $\overline{\mbox{TSE}}$, and {\em extended trajectory angular velocity} $\mbox{TRA}$,    
 $\overline{\mbox{TRA}}$. 
\changed{Single-trajectory flow measures are attractive because they need minimal information to infer the flow behavior of an underlying field. Obviously, single-trajectory measures cannot be objective in the Euclidean observation space because one may think of a reference system moving with the trajectory, making each trajectory zero \citep{Haller21singletrajecory}. Because of this, \cite{Haller21singletrajecory} considered objectivity in an extended phase space.}

Given is a $C^2$ continuous trajectory $\vx(t)$ for $t \in [t_0,t_N]$, its first and second derivatives $\dot{\vx}(t), \ddot{\vx}(t)$, and a positive constant $v_0$ accounting for a certain ratio between space and time units to make them non-dimensionalized. Considering $\vx(t)$ in an extended phase space gives for the first and second derivative of a trajectory $\underline{\vx}(t)$:
\begin{equation}
 \dot{\underline{\vx}}(t) = \begin{pmatrix}
  \frac{1}{v_0} \,   \dot{\vx}(t)\\
 1
\end{pmatrix}
\;\;\;,\;\;\;
 \ddot{\underline{\vx}}(t) = \begin{pmatrix}
\frac{1}{v_0} \, \ddot{\vx}(t)\\
 0
\end{pmatrix}.
\end{equation}
Then a local stretching measure can be defined as
\begin{equation}
\mbox{tse} = \mbox{tse}_{\vx(t),v_0}(t) = \frac{ \dot{\underline{\vx}}^\Transp \, \ddot{\underline{\vx}} }{ \dot{\underline{\vx}}^\Transp \, \dot{\underline{\vx}} }
= \frac{ \dot{\vx}^\Transp \, \ddot{\vx} }{ \dot{\vx}^\Transp \, \dot{\vx} + v_0^2}
\end{equation}
from which the Lagrangian measures $\mbox{TSE}$ and $\overline{\mbox{TSE}}$ are computed by integrating $\mbox{tse}$ along the trajectory:
\begin{align}
\label{eq_define_TSE1}
\mbox{TSE}_{\vx(t),v_0}^{t_0,t_N} &= \frac{1}{\Delta t} \int_{t_0}^{t_N} \mbox{tse} \; dt 
\;=\;
 \frac{1}{\Delta t} \ln \sqrt{  \frac{ |  \dot{\vx}(t_N)   |^2 + v_0^2  }{ |  \dot{\vx}(t_0)   |^2 + v_0^2} }
\\
\overline{\mbox{TSE}}_{\vx(t),v_0}^{t_0,t_N} &= \frac{1}{\Delta t} \int_{t_0}^{t_N} | \mbox{tse}| \, dt
\label{eq_appr_TSEbar}
\; \approx\;
 \frac{1}{\Delta t}
 \sum_{i=0}^{N-1} 
 \left|
 \ln \sqrt{  \frac{ |  \dot{\vx}(t_{i+1})   |^2 + v_0^2  }{ |  \dot{\vx}(t_i)   |^2 + v_0^2} }
 \right|
\end{align}
with $\Delta t = t_N-t_0$. The discretization in Eq.~\eqref{eq_appr_TSEbar} samples $\vx(t)$ at $N+1$ time steps $t_0 < t_1 < ... < t_N$.

For defining $\mbox{TRA}$, the $(n+1)$-dimensional matrix function
\begin{equation}
\mathbf{tra} = \mathbf{tra}_{\vx(t),v_0}(t) =
\frac{ \dot{\underline{\vx}} \,  \ddot{\underline{\vx}}^\Transp -  \ddot{\underline{\vx}} \,  \dot{\underline{\vx}}^\Transp }{  \dot{\underline{\vx}}^\Transp \,  \dot{\underline{\vx}} }
\end{equation}
can be introduced that describes the local angular velocity. Note that $\mathbf{tra}$ is an anti-symmetric matrix, from  
which one gets by integration along the trajectory Lagrangian measures
\begin{align}
\mbox{TRA}_{\vx(t),v_0}^{t_0,t_N} &= \frac{1}{\Delta t} \frac{\sqrt{2}}{2}
\left|
\int_{t_0}^{t_N}  \mathbf{tra} \; dt
\right|_{Fr}
\\
&=
\label{eq_define_TRA1}
\frac{1}{\Delta t}
\cos^{-1}
\frac{ \dot{\vx}(t_0)^\Transp  \, \dot{\vx}(t_N) + v_0^2  }{ \sqrt{  | \dot{\vx}(t_0)|^2 + v_0^2 } \sqrt{  | \dot{\vx}(t_N)|^2 + v_0^2 } }
\\
\overline{\mbox{TRA}}_{\vx(t),v_0}^{t_0,t_N} &= \frac{1}{\Delta t}  \frac{\sqrt{2}}{2}  \int_{t_0}^{t_N} \left|  \mathbf{tra} \right|_{Fr}  \, dt
\\
&\approx
\frac{1}{\Delta t}
\sum_{i=0}^{N-1}
\cos^{-1}
\frac{ \dot{\vx}(t_i)^\Transp  \, \dot{\vx}(t_{i+1}) + v_0^2  }{ \sqrt{  | \dot{\vx}(t_i)|^2 + v_0^2 } \sqrt{  | \dot{\vx}(t_{i+1})|^2 + v_0^2 } }
\end{align}
where $_{Fr}$ denotes the Frobenius norm of a matrix. \cite{Haller21singletrajecory}  claimed that  $\mbox{TSE}$ and 
$\overline{\mbox{TSE}}$ are objective in the extended phase space, and that $\mbox{TRA}$ and  $\overline{\mbox{TRA}}$ are quasi-objective in the extended phase space under a certain condition put to the average vorticity in a certain neighborhood of the trajectory.

\subsection{$\mbox{TSE}$, $\mbox{TRA}$, and underlying velocity fields}

We recapitulate the definition of $\mbox{TSE}$ from \cite{Haller21singletrajecory}, keeping their notation as much as possible. We start with a single observed trajectory
$\vx(t)$  in $n$-D ($n=2,3$) for $t \in [t_0, t_N]$ running from $\vx_0 = \vx(t_0)$ to  $\vx_N = \vx(t_N)$. Further, we assume that $\vx(t)$ is a trajectory (path line) of an underlying unsteady velocity field $\vv(\vx,t)$, i.e., $\dot{\vx}(t) = \frac{d \vx}{d t} = \vv(\vx(t),t)$ for all $t \in [t_0, t_N]$.  
Following \cite{Haller21singletrajecory}, $\vv$ is transformed into a non-dimensionalized field $\vu$ by
\begin{align}
\vy = \frac{\vx}{L}  \;\;\;,\;\;\; \tau = \tau_0 + \frac{t-t_0}{T} \;\;\;,\;\;\; v_0 = \frac{L}{T}
\label{eq:non-dimension-params}
\end{align}
where $L,T, v_0$ are certain positive constants for a field that need to be determined by additional knowledge about the data. Generally, the scaling factor $v_0$ is non-zero, i.e., $v_0\neq 0$. This transformation rephrases  $\vx(t)$ into
the non-dimensionalized trajectory 
\begin{equation}
\vy(\tau) = \frac{1}{L} \vx(t_0 + T(\tau-\tau_0))
\label{eq:define_y}
\end{equation}
running from $\vy_0=\vy(\tau_0)=\frac{1}{L} \vx_0$ to $\vy_N=\vy(\tau_N)=\frac{1}{L} \vx_N$ with $\tau_N = \tau_0 + \frac{t_N-t_0}{T}$.
Further, it gives the non-dimensionalized vector field
\begin{equation}
\label{eq_define_u}
\vu(  \vy, \tau) =  \frac{1}{v_0} \vv \left( L \vy , t_0 + T (\tau- \tau_0) \right).
\end{equation}
Note that (\ref{eq_define_u}) contains a correction of a missing term $\frac{1}{v_0}$
 in formula (26) in \citep{Haller21singletrajecory}. The error in  formula (26) in \changed{\citep{Haller21singletrajecory}} can be seen in the following way: suppose $\vv$ is a constant vector field, i.e., $\vv(\vx,t)=\vv_c$. Then formula (26) in \citep{Haller21singletrajecory} would give $\vu(\vy,\tau)=\vv_c$ no matter how $v_0$ is chosen. This would contradict to the formula before (33) in \citep{Haller21singletrajecory}.

Following \cite{Haller21singletrajecory} further, an extended phase space $\mY = 
\begin{pmatrix}
  \vy\\
 z
\end{pmatrix}
$ is introduced \changed{by adding the additional time dimension $z$}. Transformation of $\vy(\tau)$ and $\vu(  \vy, \tau)$ into this extended phase space gives
\begin{equation}
\mY(\tau) = 
\begin{pmatrix}
  \vy(\tau)\\
 \tau
\end{pmatrix}
\;\;\;,\;\;\; 
\mU(\mY) = 
\begin{pmatrix}
  \vu(\vy,z)\\
 1
\end{pmatrix}
\label{eq:define-extended-phase-space}
\end{equation}
where $\mY(\tau)$ is the trajectory in the extended phase space 
running from $\mY_0 = \mY(\tau_0) = 
 \begin{pmatrix}
   \vy_0\\
 \tau_0
\end{pmatrix}$ to  $\mY_N = \mY(\tau_N) = 
 \begin{pmatrix}
   \vy_N\\
 \tau_N
\end{pmatrix}$, and $\mU(\mY)$ is the underlying vector field. The tangent vector of $\mY(\tau)$ is
\begin{equation}
\mY'(\tau) = \frac{ d \, \mY}{ d \, \tau} =
 \begin{pmatrix}
   \vy'(\tau)\\
1
\end{pmatrix}
=
\begin{pmatrix}
 \frac{1}{v_0}  \dot{\vx}( t_0 + T(\tau-\tau_0 ) )\\
1
\end{pmatrix}.
\end{equation}
Note that $\mU(\mY)$ is an autonomous dynamical system now: $\mU$ is  a steady velocity field in the extended phase space. Then Haller et al.~\cite{Haller21singletrajecory} defines $\mbox{TSE}$ and $\mbox{TRA}$ as
 \begin{eqnarray}
\label{eq_define_TSE2} 
\mbox{TSE}_{t_0}^{t_N}(\vx_0,v_0) &=& 
 \frac{1}{\Delta t} \ln \frac{| \mY'(\tau_N) | }{ | \mY'(\tau_0) | }
= \frac{1}{\Delta t} \ln \frac{| \mU(\mY_N) | }{ | \mU(\mY_0) | }
\\
\label{eq_define_TRA2} 
\mbox{TRA}_{t_0}^{t_N}(\vx_0,v_0) &=& 
 \frac{1}{\Delta t} \cos^{-1} \frac{ \mY'(\tau_0)^\Transp \, \mY'(\tau_N)   }{ | \mY'(\tau_0) | |\mY'(\tau_N)| }
 \\
&=& \frac{1}{\Delta t} \cos^{-1} \frac{\mU(\mY_0)^\Transp \, \mU(\mY_N)  }{ | \mU(\mY_0) | | \mU(\mY_N) | }
\end{eqnarray}
where $\Delta t = t_N-t_0$, (\ref{eq_define_TSE2}) is identical to the right-hand side of (\ref{eq_define_TSE1}),
and (\ref{eq_define_TRA2}) is identical to the right-hand side of (\ref{eq_define_TRA1}).
To show objectivity of $\mbox{TSE}$ in the extended phase space, one has to prove that $\mbox{TSE}$ is invariant under observation in any moving Euclidean reference system in the extended phase space. Analogous to Eq.~\eqref{eq_define_movingsystem}, such moving reference system is defined by
 \begin{equation}
 \label{eq_define_moving_frame_1}
 \mY = 
 \boldsymbol{\mathcal Q}(\tau)  \widetilde{\mY} + \mB(\tau)
 \;,\;
   \boldsymbol{\mathcal Q}(\tau)
 =
 \begin{pmatrix}
\mQ(\tau) &  \vNull \\
 \vNull^\Transp & 1
\end{pmatrix}
\;,\;
\mB(\tau) =
 \begin{pmatrix}
\vb(\tau)  \\
0
\end{pmatrix}
\end{equation}
with $\mQ(\tau) \in SO(n)$ being a rotation matrix, and $\vNull$ being the zero-vector. The observed trajectory $\widetilde{\mY}(\tau)$ and the underlying velocity field  $\widetilde{\mU}(\widetilde{\mY},\tau)$ in the new moving reference system are 
 \begin{align}
\label{eq_trafo_in_extended_reference_system2} 
\widetilde{\mY}(\tau) &=     \boldsymbol{\mathcal Q}^\Transp(\tau)  (\mY(\tau) - \mB(\tau))
\\
\label{eq_trafo_in_extended_reference_system}
\widetilde{\mU}(\widetilde{\mY},\tau) &=   \boldsymbol{\mathcal Q}^\Transp(\tau)  
\left(
\mU
\left(    \boldsymbol{\mathcal Q}(\tau) \widetilde{\mY} + \mB(\tau)  \right) - \dot{  \boldsymbol{\mathcal Q}}(\tau)  \widetilde{\mY} -  \dot{\mB}(\tau) 
\right)
 \end{align}
where the new trajectory $\widetilde{\mY}(\tau)$ runs from $\widetilde{\mY}_0 = \widetilde{\mY}(\tau_0)$ to 
$\widetilde{\mY}_N = \widetilde{\mY}(\tau_N)$. Then,  $\mbox{TSE}$ in the moving reference system is
 \begin{eqnarray}
 \label{eq_define_TSEtilde}
\widetilde{\mbox{TSE}}_{t_0}^{t_N}(\vx_0,v_0) &=& 
\frac{1}{\Delta t} \ln \frac{ |  \widetilde{\mY}'(\tau_N)   | }{ | \widetilde{\mY}'(\tau_0) |  }
=
\frac{1}{\Delta t} \ln \frac{ | \widetilde{\mU}(\widetilde{\mY}_N, \tau_N) | }{ | \widetilde{\mU}(\widetilde{\mY}_0, \tau_0) |  }.
\\
 \label{eq_define_TRAtilde}
\widetilde{\mbox{TRA}}_{t_0}^{t_N}(\vx_0,v_0) &=& 
\frac{1}{\Delta t} \cos^{-1} \frac{ \widetilde{\mY}'(\tau_0)^\Transp \,  \widetilde{\mY}'(\tau_N)   }{ | \widetilde{\mY}'(\tau_0) | |  \widetilde{\mY}'(\tau_N)   |  }
 \end{eqnarray}
 To show objectivity of $\mbox{TSE}$ in the extended phase space, one has to prove $\mbox{TSE} = \widetilde{\mbox{TSE}}$ for any moving reference frame, as given by Eq.~\eqref{eq_define_moving_frame_1}.
To show quasi-objectivity of $\mbox{TRA}$ under averaged-vorticity condition, one has to prove  
$\mbox{TRA} = \widetilde{\mbox{TRA}}$ for all reference frames (\ref{eq_define_moving_frame_1}) in which the averaged-vorticity condition is fullfilled.

\section{The first counter-example}
\label{sec_counterexample}

We show the non-objectivity of $\mbox{TSE}$ in the extended phase space by a simple counter-example. We set the 2D observed trajectory $\vx(t)$ and the underlying
velocity field $\vv(\vx,t)$ as
\begin{equation}
\label{eq_definetrajectory}
\vx(t) = 
\begin{pmatrix}
  e^t - 1 \\
   t \, (t+1)
\end{pmatrix}
\;\;\;,\;\;\;
\vv(\vx,t) = 
\begin{pmatrix}
  x+1 \\
   2 \, t+1
\end{pmatrix}
\end{equation}
for $t \in [t_0, t_N]=[0,1]$   and  $\vx = (x,y)^\Transp$. To calculate $\mbox{TSE}$ as in Eq.~\eqref{eq_define_TSE1}, we only need information at time $t_0$ and $t_N$. This gives
\begin{eqnarray}
\vx_0 = 
(0,0)^\Transp
&,&
\vx_N = 
(e-1,2)^\Transp
\\
\dot{\vx}(t_0) = \vv( \vx_0, t_0) = 
(1,1)^\Transp
&,&
\dot{\vx}(t_N) = \vv( \vx_N, t_N) = 
(e,3)^\Transp.
\end{eqnarray}
For the non-dimensionalization transformation, we set $\tau_0=0$, resulting in $\tau_N= \frac{1}{T}$. This gives with Eqs.~\eqref{eq:define_y} and \eqref{eq_define_u}
\begin{equation}
\vy(\tau) = \frac{1}{L}
\begin{pmatrix}
  e^{T \tau}  - 1\\
   T \tau (T \tau+1)
\end{pmatrix}
\;\;\;,\;\;\;
\vu(  \vy, \tau) = 
\frac{1}{v_0}  \begin{pmatrix}
 L  \bar{x}+1 \\
   2 \, T \tau+1
\end{pmatrix}
\end{equation}
with  $\vy = (\bar{x}, \bar{y})^\Transp$, and therefore we obtain at $\tau_0$ and $\tau_N$
\begin{eqnarray}
\vy_0  = 
\begin{pmatrix}
  0 \\
   0
\end{pmatrix}
&,&
\vy_N = \frac{1}{L}  
\begin{pmatrix}
  e-1 \\
   2
\end{pmatrix}
\\
\vu(\vy_0,\tau_0) =\frac{1}{v_0}
 \begin{pmatrix}
  1 \\
   1
\end{pmatrix}
&,&
\vu(\vy_N,\tau_N) =\frac{1}{v_0}
 \begin{pmatrix}
  e \\
   3
\end{pmatrix}.
\end{eqnarray}
Transforming to the extended phase space 
\begin{equation}
\mY = ( \bar{x}, \bar{y}, z )^\Transp
\end{equation}
using Eq.~\eqref{eq:define-extended-phase-space} gives 
 \begin{equation}
\mY(\tau) = 
\begin{pmatrix}
  \frac{1}{L} (e^{T \tau}  - 1)\\
   \frac{1}{L} ( T \tau (T \tau+1)) \\
   \tau
\end{pmatrix}
\;\;\;,\;\;\;
\mU(\mY) = 
\begin{pmatrix}
\frac{1}{v_0} ( L  \bar{x}+1) \\
\frac{1}{v_0} ( 2 \, T z +1)\\
1
\end{pmatrix}
\end{equation}
with the following position and tangent at the curve end points
\begin{eqnarray}
\mY_0  = 
(0,0,0)^\Transp
&,&
\mY_N =
\left(\frac{e-1}{L} , \frac{2}{L} , \frac{1}{T} \right)^\Transp
\\
\mU(\mY_0) =
\left(\frac{1}{v_0} , \frac{1}{v_0} , 1\right)^\Transp
&,&
\mU(\mY_N) =
\left(\frac{e}{v_0} , \frac{3}{v_0} , 1 \right)^\Transp.
\end{eqnarray}
Inserting into Eqs.~\eqref{eq_define_TSE2} and \eqref{eq_define_TRA2}, this results in \mbox{TSE} and \mbox{TRA}:
\begin{equation}
\label{eq_define_tse2}
\mbox{TSE} = \ln \sqrt{ \frac{ e^2 + 9 + v_0^2 }{ 2 + v_0^2 }  }
\;\;,\;\;
\mbox{TRA} = \cos{-1} \frac{ e + 3 + v_0^2 }{ \sqrt{2 + v_0^2}\sqrt{e^2 + 9 + v_0^2}  }.  
\end{equation}
For our counterexample, it is sufficient to choose a particular moving Euclidean reference system
 (\ref{eq_define_moving_frame_1}) by
 \begin{equation}
 \label{eq_Galilean_moving_frame}
   \boldsymbol{\mathcal Q}(\tau) = \mI \;\;\;,\;\;\; \mB(\tau) = \tau 
  \begin{pmatrix}
\vb_c \\
0
\end{pmatrix}
 \end{equation}
where $\mI$ is the identity matrix and $\vb_c = (x_c,y_c)^\Transp$ is a constant 2D vector. For this particular reference system, we get by 
(\ref{eq_trafo_in_extended_reference_system2}),
(\ref{eq_trafo_in_extended_reference_system}):
 \begin{eqnarray}
\widetilde{\mY}(\tau) &=& \mY(\tau) - \tau  \begin{pmatrix}
\vb_c \\
0
\end{pmatrix}
\\
\widetilde{\mU}(\widetilde{\mY},\tau) &=& \mU \left( \mY + \tau
\begin{pmatrix}
\vb_c \\
0
\end{pmatrix}
\right) 
-
\begin{pmatrix}
\vb_c \\
0
\end{pmatrix}.
\end{eqnarray} 
 This gives the following trajectory end points and tangents:
\begin{eqnarray}
\widetilde{\mY}_0  = 
(0,0,0)^\Transp
\;\;\;  ,\;\;\;
\widetilde{\mY}_N =
\left( \frac{e-1}{L}  -  \frac{x_c}{T} \,\,,\,\,
\frac{2}{L}  -  \frac{y_c}{T} \,\,,\,\, \frac{1}{T} \right)^\Transp
\\
\label{eq_YTilde0}
\widetilde{\mU}(\widetilde{\mY}_0,\tau_0) =
\widetilde{\mY}'(\tau_0)
=
\left( \frac{1}{v_0}  - x_c \,\,,\,\,
\frac{1}{v_0}  - y_c \,\,,\,\, 1 \right)^\Transp
\\
\label{eq_YTildeN}
\widetilde{\mU}(\widetilde{\mY}_N, \tau_N) =
 \widetilde{\mY}'(\tau_N)
=
\left( \frac{e}{v_0}  - x_c \,\,,\,\,
\frac{3}{v_0} - y_c \,\,,\,\, 1 \right)^\Transp
\end{eqnarray}
and finally by inserting into Eq.~\eqref{eq_define_TSEtilde}, we get $\widetilde{\mbox{TSE}}$:
\begin{equation}
\label{eq_tilde_tse}
\widetilde{\mbox{TSE}} = \ln \sqrt{ \frac{ (e-v_0 x_c)^2 + (3 - v_0 y_c)^2 + v_0^2 }{  (1 - v_0 x_c)^2 +  (1 - v_0 y_c)^2 + v_0^2 }  }
\end{equation}
Analogously, $\widetilde{\mbox{TRA}}$ follows by inserting (\ref{eq_YTilde0})--(\ref{eq_YTildeN}) into (\ref{eq_define_TRAtilde}).
Since there is no positive constant $v_0$, cf. \eqref{eq:non-dimension-params}, that makes $\mbox{TSE}$   in (\ref{eq_define_tse2})   and   $\widetilde{\mbox{TSE}} $ in (\ref{eq_tilde_tse}) identical for any 
$\vb_c = (x_c,y_c)^\Transp$, non-objectivity of $\mbox{TSE}$ in the extended phase space is shown.
Since in our example both $\mU(\mY)$ and
$\widetilde{\mU}(\widetilde{\mY},\tau)$ have zero vorticity, the average-vorticity condition in  
Haller et al.~\citep{Haller21singletrajecory} is trivially fulfilled. Thus, the difference of $\mbox{TRA}$ and  $\widetilde{\mbox{TRA}}$ gives that $\mbox{TRA}$ is not quasi-objective in the extended phase under the average-vorticity condition.

\subsection*{Where is the error?}

Haller et al.~\cite{Haller21singletrajecory} considered a non-zero vector $\vxi_0$ at $(\vx_0,t_0)$ that is advected with $\vv$ along $\vx(t)$, resulting in
\begin{equation}
\label{eq_pde_general}
\dot{\vxi}(t) = \nabla \vv(\vx(t),t) \; \vxi(t) 
\;\;\; ,\; \; \;
 \vxi(t_0) =   \vxi_0.
\end{equation}
Then, $\vxi(t)$ is observed under a moving reference system (\ref{eq_define_movingsystem}).
Objectivity of $\vxi$ is deduced from  (\ref{eq_pde_general}), (\ref{eq_define_movingsystem}):
\begin{equation}
\label{eq_defineobjectovpy_xi}
\widetilde{ \vxi }(t) =  \mQ^\Transp(t)  \; \vxi(t) 
\end{equation}
where $\widetilde{ \vxi }$ is the observation of $\vxi$ under the moving reference system (\ref{eq_define_movingsystem}). From 
(\ref{eq_defineobjectovpy_xi}) follows the objectivity of 
$\frac{1}{\Delta t} \ln \frac{| \vxi(t_N) | }{| \vxi_0 |}$.
We note that (\ref{eq_defineobjectovpy_xi}) follows from (\ref{eq_pde_general}) and (\ref{eq_define_movingsystem}) only if another implicit assumption holds: objectivity of the seeding vector  $\vxi_0$, i.e.,  $\widetilde{ \vxi }_0 =  \mQ^\Transp(t_0)  \;   \vxi_0$.

The approach of Haller et al.~\cite{Haller21singletrajecory} is to set $\vxi_0 = \vv_0 = \vv(\vx_0, t_0)$. With this, additional conditions are  necessary  to ensure
\begin{eqnarray}
\label{eq_cond_obj_1}
\dot{\vv}(t) &=& \nabla \vv(\vx(t),t) \; \vv( \vx, t) 
\\
\label{eq_cond_obj_2}
\widetilde{\vv}( \widetilde{\vx} , t) &=&  \mQ^\Transp(t)  \; \vv( \vx , t) 
\end{eqnarray}
where (\ref{eq_cond_obj_1}) corresponds to  (\ref{eq_pde_general}) and (\ref{eq_cond_obj_2}) corresponds to  (\ref{eq_defineobjectovpy_xi}).
 To ensure (\ref{eq_cond_obj_1}), Haller et al.~\cite{Haller21singletrajecory} introduced the condition
\begin{equation}
\nonumber
\mbox{(A1)}  \;\;\;\;\;\;  \;\;\;\;\;\; \;\;\;\;\;\;  \;\;\;\;\;\; 
 \;\;\;\;\;\;  \partial_t \vv(\vx,t) = \vNull
\end{equation}
in the current observation frame. However, condition (A1) does not ensure  (\ref{eq_cond_obj_2}) because $\vxi_0=\vv_0$ is not objective. Since the observation of 
$\vv$ under the moving reference system (\ref{eq_define_movingsystem}) is \citep{Haller20:can}
\begin{equation}
\widetilde{\vv}(\widetilde{\vx},t) = \mQ^\Transp(t) \left(  \vv(\vx,t)  - \dot{\mQ}(t) \; \widetilde{\vx} - \dot{\vb}(t) \right),
\end{equation}
Eq.~\eqref{eq_cond_obj_2} is \changed{in general} only fulfilled for $ \dot{\mQ}=\vNull, \dot{\vb}=\vNull$, i.e., the reference frame is not moving but static, resulting in demanding that $\widetilde{\vv}(\widetilde{\vx},t)$ is steady. This means that the condition for the quasi-objectivity of TSE is the steadiness of both 
$\vv$ and $\widetilde{\vv} $ in all considered reference frames. We remark that this is a rather strong condition for quasi-objectivity: it excludes the consideration of all moving reference frames.

The transformation to the extended reference system transforms $\vv$ to the steady vector field $\mU$, making the condition (A1) for (\ref{eq_cond_obj_1}) in the extended reference frame obsolete. However, the observation  $\widetilde{\mU}$  of $\mU$ under a moving reference system
 (\ref{eq_define_moving_frame_1}) is \emph{not} a steady vector field anymore, as shown in
 (\ref{eq_trafo_in_extended_reference_system}). This means that
 \begin{equation}
  \widetilde{\mU}( \widetilde{\mY} , \tau ) =
   \boldsymbol{\mathcal Q}(\tau) \;
   \mU(\mY) 
 \end{equation} 
 does not hold in general but only for particular steady reference frames. Because of this, TSE is in the extended phase space not objective but only quasi-objective under restriction to a static reference system.
 
\paragraph*{Summary:}
The error was to assume that the observation of an autonomous system (steady vector field) in the extended phase space under a moving reference frame remains an autonomous system.

\paragraph*{Remarks:}
A similar argumentation gives that $\overline{\mbox{TSE}} $ is not objective in the extended phase space, and and that
$\overline{\mbox{TRA}}$ is is not quasi-objective in the extended phase space under the averaged-vorticity-based condition.
$\mbox{TSE}$, $\overline{\mbox{TSE}} $, $\mbox{TRA}$ and $\overline{\mbox{TRA}}$ are not even Galilean invariant because the moving reference system 
 (\ref{eq_Galilean_moving_frame}) in the counterexample was performing a Galilean transformation.

\section{The second counter-example}

In an erratum, Haller et al.~\cite{Haller21singletrajecory_erratum} give up the idea of considering extended phase space and  non-dimensionalization, and  introduce a new condition 
\begin{equation}
 (A3)  \; \;\;\;\;\;\;\;\;\; \;\;\;\;\;\;\;\;
|  \partial_t \vv(\xx(t),t) | \ll | \ddot{\xx}(t)  |.
\end{equation}
Based on this, Theorem 3 in \cite{Haller21singletrajecory} is reformulated. Unfortunately, the reformulated theorem is still incorrect.
To show this, we consider a second counter-example. We consider the 3D trajectory and the underlying velocity field
\begin{equation}
\label{eq_example2}
\vx(t) = \begin{pmatrix}
 e^t\\-t\\0 
\end{pmatrix}
\;\;\; , \;\;\;
\vv(\vx,t) = 
\begin{pmatrix}
 x\\-1\\0 
\end{pmatrix}
\end{equation}
with $\vx=(x,y,z)^\Transp$ and $t$ running from $t_0=0$ to $t_N=1$. Observing $\vx, \vv$ under a reference system 
(\ref{eq_define_movingsystem}) with 
\begin{equation}
\label{eq_example2_referencesystem}
\mQ(t)=\mI  \;\;\;,\;\;\;   \vb(t) = (0,t,0)^\Transp 
\end{equation}
gives
\begin{equation}
\label{eq_example2_moved}
\widetilde{\vx}(t) = \begin{pmatrix}
 e^t\\- 2 \; t\\0 
\end{pmatrix}
\;\;\; , \;\;\;
\widetilde{\vv}(\widetilde{\vx},t) = 
\begin{pmatrix}
 \widetilde{x}\\-2\\0 
\end{pmatrix}
\end{equation}
with $\widetilde{\vx}=(\widetilde{x},\widetilde{y},\widetilde{z})^\Transp$. We note that \begin{eqnarray}
\partial_t \vv(\vx,t) = \partial_t \widetilde{\vv}(\widetilde{\vx},t) = 
\vomega(\vx,t) = \widetilde{\vomega}(\widetilde{\vx},t) =
\vNull \label{eq:vtzero}
\\
\ddot{\xx}(t) = \widetilde{\ddot{\xx}}(t) = (e^t,0,0)^\Transp \label{eq:anonzero}
\end{eqnarray}
where $\vomega, \widetilde{\vomega}$ are the vorticity of $\vv, \widetilde{\vv}$, respectively. This means that all conditions (A1),(A2) (from \cite{Haller21singletrajecory}) and (A3) (from \cite{Haller21singletrajecory_erratum}) are fulfilled both in the original frame in (\ref{eq_example2}) and in the particular moving reference frame in (\ref{eq_example2_moved}). Computing $\mbox{TSE}$, 
$\overline{\mbox{TSE}}$, $\mbox{TRA}$, 
$\overline{\mbox{TRA}}$ from (\ref{eq_example2}), and computing  
$\widetilde{\mbox{TSE}}$, 
$\widetilde{\overline{\mbox{TSE}}}$, $\widetilde{\mbox{TRA}}$, 
$\widetilde{\overline{\mbox{TRA}}}$ from (\ref{eq_example2_moved}) 
gives
\begin{align}
\label{eq_counterexample_1}
\mbox{TSE} = \overline{\mbox{TSE}}&=\frac{\ln(e^2+1)-\ln(2)}{2}  &\approx 0.71689\\
\label{eq_counterexample_2}
\widetilde{\mbox{TSE}} = \widetilde{\overline{\mbox{TSE}}} &= \frac{\ln(e^2+4)-\ln(5)}{2}  &\approx 0.41161\\
\mbox{TRA} = \overline{\mbox{TRA}} &= \tan^{-1}(e)-\frac{\pi}{4}  &\approx 0.43288\\
\label{eq_counterexample_4}
\widetilde{\mbox{TRA}} = \widetilde{\overline{\mbox{TRA}}} &= \tan^{-1}(e/2)-\tan^{-1}(1/2)  &\approx 0.47282
\end{align}
\changed{For $\mbox{TSE}$ being quasi-objective under condition (A3) following \cite{Haller21singletrajecory}, there must be an objective scalar $\cP$ such that
\begin{equation}
\label{eq_TSE_conterexample2}
\mbox{TSE} \approx \cP \;\;\;,\;\;\;  \widetilde{\mbox{TSE}} \approx \widetilde{\cP}
\end{equation}
where $\widetilde{\cP}$ is the observation of $\cP$ under reference frame (\ref{eq_example2_referencesystem})
and
``the accuracy of the approximation indicated by the symbol $\approx$ depends on the extend to which the conditions (A3) is satisfied''~\cite{Haller21singletrajecory}.
Note that here assumption (A3) is completely satisfied, since the value 
$|  \partial_t \vv(\xx(t),t) | / | \ddot{\xx}(t)  |$ as considered in Figure 1 of 
 \cite{Haller21singletrajecory_erratum} is constant zero in both reference frames, see Eq.~\eqref{eq:vtzero}--\eqref{eq:anonzero}.
Since $\cP=\widetilde{\cP}$ due to the demanded objectivity of $\cP$ and since assumption (A3) is completely satisfied, 
Eqs. (\ref{eq_counterexample_1}) and (\ref{eq_counterexample_2}) give that 
(\ref{eq_TSE_conterexample2}) cannot be fulfilled since $\mbox{TSE}$ and $\widetilde{\mbox{TSE}}$ are significantly different. Hence, $\mbox{TSE}$ is not quasi-objective under condition (A3). }

\changed{Eqs.~(\ref{eq_counterexample_1})--(\ref{eq_counterexample_4}) give that similar statements hold for all measures $\mbox{TSE}$, 
$\overline{\mbox{TSE}}$, $\mbox{TRA}$, 
$\overline{\mbox{TRA}}$: none of them is quasi-objective under any combination of conditions (A1), (A2), (A3).}
This example and the example from Section 
\ref{sec_counterexample}
show that all theorems in both \cite{Haller21singletrajecory} and 
\cite{Haller21singletrajecory_erratum} are incorrect.

\bibliographystyle{unsrtnat}
\bibliography{literature}  






\end{document}